# Phenolic Resin Dual-Use Stamps for Capillary Stamping and Decal Transfer Printing

Leiming Guo,[†,*] Jonas Klein,[†] Jannis Thien,[‡] Michael Philippi,[†] Markus Haase,[†] Joachim Wollschläger,[‡] and Martin Steinhart[†,*]

[†] Institut für Chemie neuer Materialien and CellNanOs, Universität Osnabrück, Barbarastr. 7, 49076 Osnabrück, Germany

[‡] Department of Physics, Universität Osnabrück, Barbarastr. 7, 49076 Osnabrück, Germany

**ABSTRACT**

We report an optimized two-step thermopolymerization process carried out in contact with micropatterned molds that yields porous phenolic resin dual-use stamps with topographically micropatterned contact surfaces. With these stamps, two different parallel additive substrate manufacturing methods can be executed: capillary stamping and decal transfer microlithography. Under moderate contact pressures, the porous phenolic resin stamps are used for non-destructive ink transfer to substrates by capillary stamping. Continuous ink supply through the pore systems to the contact surfaces of the porous phenolic resin stamps enables multiple successive stamp-substrate contacts for lithographic ink deposition under ambient conditions. No deterioration of the quality of the deposited pattern occurs and no interruptions for ink replenishment are required. Under high contact pressure, porous phenolic resin stamps are used for decal transfer printing. In this way, the tips of the stamps' contact elements are lithographically transferred to counterpart substrates. The granular nature of the phenolic resin facilitates the rupture of the contact elements upon stamp retraction. The deposited phenolic resin micropatterns characterized by abundance of exposed hydroxyl groups are used as generic anchoring sites for further application-specific functionalizations. As example, we deposited phenolic resin micropatterns on quartz crystal microbalance resonators and further functionalized them with polyethylenimine for preconcentration sensing of humidity and gaseous formic acid. We envision that also preconcentration coatings for other sensing methods, such as attenuated total reflection infrared spectroscopy and surface plasmon resonance spectroscopy, are accessible by this functionalization algorithm.

# 1. Introduction

Additive substrate manufacturing[1] involving the deposition of functional materials on counterpart substrates is ubiquitous in academic research and industry-scale production. A process comprising the simultaneous large-scale generation of patterns of a deposited material in extended substrate areas by means of a suitable lithographic method is called parallel and distinguished from serial patterning processes involving pixel-by-pixel writing. Parallel additive substrate manufacturing is attractive because of its high throughput. In particular, contact lithography with micropatterned stamps has been explored as parallel additive substrate manufacturing technique to produce components with tailored properties for electronics, optics, heat management, biomedical analytics, sensing and many more applications. A first relevant embodiment of parallel additive contact lithography includes the transfer of ink from a stamp to a substrate by classical microcontact printing[2-6] and polymer pen lithography[7-9] using solid elastomeric stamps as well as by capillary stamping using porous homopolymer,[10] block copolymer[11-13] and silica[14, 15] stamps. A second embodiment of parallel additive contact lithography is decal transfer microlithography, where parts of a stamp are lithographically transferred to a substrate.[16-21] The stamps are pressed against the substrate so that the stamps' contact elements form tight contact to the substrates. Upon stamp retraction, the contact elements rupture so that their tips remain attached to the substrates.

Phenolic resins[22-24] are crosslinked thermosets chemically stable against organic solvents as well as acids and bases, which are synthesized by polycondensation of phenols and aldehydes. Water is a by-product of these polycondensation reactions. The volumes occupied by the water are converted into pores. Hence, the presence of pores is an intrinsic feature of phenolic resins.[25] Phenolic resins even contain voids when prepared under compressive stress.[26] Parameswaran et al. state that the "…main disadvantages of phenolic resin are microvoid formation in the cured resin and the brittle nature resulting from a high degree of crosslinking".[27] Examples for the optimization of synthetic strategies to enhance the usability of porous phenolic resins include the formation of network-reinforcing molecular bridges by Friedel–Crafts alkylations[28] and polymerization under hypersaline conditions.[29] Thus, porous phenolic resins were used for carbon capture[28] and as material for separation membranes.[23, 30, 31]

Here we report an optimized two-step thermopolymerization combined with replication molding, which yields porous phenolic resin dual-use stamps with a granular-porous morphology and with topographically micropatterned contact surfaces for parallel additive microcontact lithography. Under moderate contact pressures (here about 5.3 kN m$^{-2}$), these porous phenolic resin stamps are used under ambient conditions for manual non-destructive capillary stamping of dye solutions selected as model inks. Thus, dye microdot arrays were formed on counterpart substrates by a multitude of successive lithographic stamp-substrate contacts. No loss of pattern quality occurs and no interruptions for ink replenishment are required because ink can be supplied to the contact surfaces of the stamps anytime during stamping through the interstices between the phenolic resin nanoparticles the stamps consist of. The second use of the porous phenolic resin stamps is decal transfer printing, which we studied for the following reasons. Phenolic resin is characterized by abundance of exposed hydroxyl groups. Therefore, phenolic resins are of interest as components in coatings capturing analytes for preconcentration sensing[32-35] by, for example, attenuated total reflection (ATR) infrared spectroscopy, surface plasmon resonance (SPR) spectroscopy or quartz crystal microbalance (QCM) sensing.[36] Analyte molecules may interact directly with the hydroxyl groups of the phenolic resin, or the hydroxyl groups of the phenolic resin are used as binding sites for further application-specific functionalizations. Colloidal phenolic resin nanoparticles can be synthesized by an adaptation of the Stöber method.[37] However, additive substrate manufacturing to realize sensor components consisting of phenolic resin microstructures tightly adhering to functional substrates, such as QCM resonators, has remained challenging. In general, composite systems comprising a continuous coating on a counterpart substrate may suffer from poor adhesion, delamination and the formation of wrinkles. These drawbacks are amplified if the coatings are brittle,[38-40] if the substrate surfaces are rough or possess step edges and if dust particles are present.[41] In this context, phenolic resin coatings are particularly problematic because phenolic resins are brittle and significantly shrink during curing.[27, 42, 43] We did thus not succeed in the direct synthesis of useable phenolic resin layers on functional substrates. Instead, we have studied decal transfer printing of phenolic resin microdots on QCM resonators as viable alternative having the following advantages. 1) The high contact pressure (here about 2368 kN m$^{-2}$) ensures strong adhesion of the transferred parts of the porous phenolic resin stamps to the counterpart substrate, whereas the granular nature of the phenolic resin resulting from the two-step thermopolymerization process introduced here facilitates the rupture of the contact elements upon stamp retraction. 2) The micropattern consisting of discrete microdots is resistant against delamination and wrinkling because delamination and wrinkling events remain confined to single affected phenolic resin microdots; the uncoated (exposed) substrate areas between the phenolic resin microdots stop delamination and wrinkling. 3) The phenolic resin microdots may capture analyte molecules themselves, or they act as generic anchoring sites for application-specific post-stamping functionalizations. 4) Functionalizations of the phenolic resin microdots with long-chain functional polymers for the capturing of analyte molecules, such as polyethylenimine (PEI) with a large number of exposed amino groups, may increase the microdot/substrate interface significantly, as discussed below. Segments of the functional polymer chains immobilized on the phenolic resin microdots will also directly be adsorbed on the substrates. Even if only the long-chain functional polymers capture analyte molecules, significant amounts of analyte molecules will thus be located in the immediate proximity of the substrate surface. We report the exemplary functionalization of QCM resonators with phenolic resin microdots, onto which in turn PEI is adsorbed, and demonstrate efficient preconcentration sensing of humidity and gaseous formic acid with thus-functionalized QCM resonators.

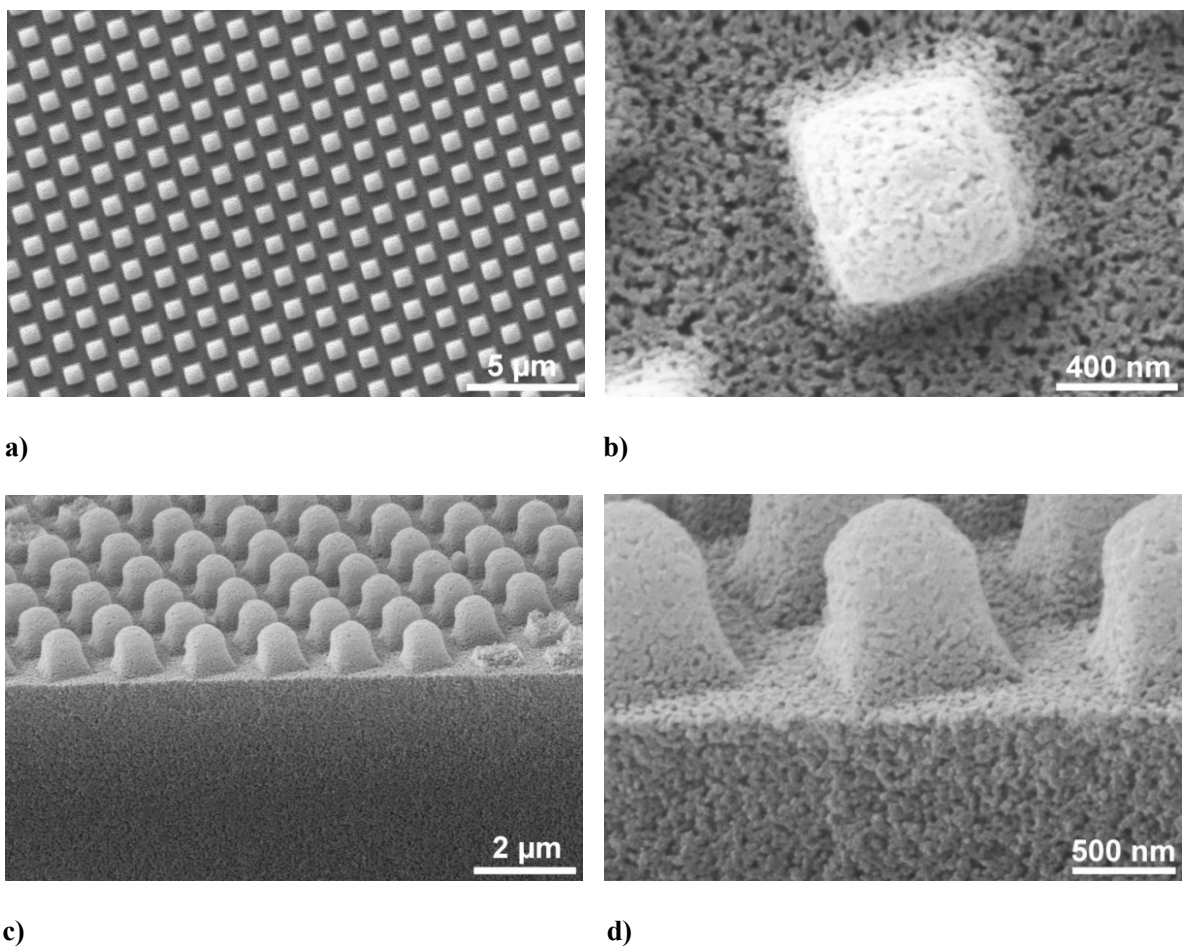

a)

b)

c)

d)

**Figure 1.** Scanning electron microscopy (SEM) images of a porous phenolic resin stamp fabricated by two-step thermopolymerization. a, b) Top views; c, d) cross sectional views.

## 2. Results and discussion

**2.1 Preparation of phenolic resin stamps.** We prepared porous phenolic resin stamps with topographically patterned contact surfaces using the same macroporous silicon molds[44] as in a previous work.[11] These macroporous silicon molds contained hexagonal arrays of macropores with a lattice constant of 1.5 µm (Figure S1). The tapered macropores with hemispherical macropore bottoms had a depth of ≈810 nm and a base diameter of ≈1 µm. We developed a two-step thermopolymerization protocol compatible with replication molding against macroporous silicon (Figure S2), which is based on a previously reported synthetic procedure involving the simultaneous gelation of poloxamers and thermopolymerization of resol oligomers in the presence of $ZnCl_2$ at elevated temperatures.[23, 29-31] An ethanolic solution containing resol, the poloxamer P123 ($PEO_{20}$-$b$-$PPO_{70}$-$b$-$PEO_{20}$; PEO = poly(ethylene glycol); PPO = poly(propylene glycol)) and $ZnCl_2$ was drop-cast onto the macroporous silicon molds. The samples were then kept at 40°C for 2h under ambient conditions. This first annealing step at 40°C helps prevent undesired coarsening of the morphology of the phenolic resin. Under these conditions, the resol mixes with the PEO block segments and is repelled from the PPO block segments of P123 when the ethanol evaporates (Figure S3a, b). The $ZnCl_2$ may coordinate to resol molecules and to ethylene oxide repeat units of the P123 and acts, therefore, as a linker between P123 and resol. The morphology obtained in this way is arrested by the inset of moderate thermopolymerization crosslinking adjacent resol molecules (Figure S3c), which is accelerated by the $ZnCl_2$.[23, 29] The $ZnCl_2$ anchors the PEO block segments of the P123 in the formed phenolic resin particles with diameters of ≈30 nm to ≈40 nm. P123 molecules having their PEO segments anchored in adjacent phenolic resin particles in turn stabilize the obtained granular phenolic resin. Further thermopolymerization at 100 °C for 12 h results in structural reinforcement by formation of additional bonds between the phenolic resin particles (Figure S3d). To enhance the porosity of the formed phenolic resin, the poloxamer and the Zn compounds were removed by treatment with concentrated sulfuric acid at elevated temperatures following a procedure reported by Zhuang et al.[45] As a result, the granular structure of the phenolic resin particles is penetrated by continuous mesopore networks with mesopore diameters ranging from ≈10 nm to ≈30 nm. The porous phenolic resin stamps obtained in this way could readily be peeled off from the macroporous silicon after the sulfuric acid treatment (Figure S2c) and were then transferred onto silicon substrates (Figure S2d). The recovered macroporous silicon molds (Figure S4) are recyclable and can be reused.

The synthetic procedure devised here was optimized so as to obtain phenolic resin with a granular structure that can be used as stamp material for capillary stamping and decal transfer microlithography. The omission of the first thermopolymerization step and direct heating of the reaction mixture to 100°C resulted in rapid thermopolymerization and consequently the formation of large phenolic resin aggregates with diameters well above 150 nm (Figure S5a).[23] Phenolic resins prepared by molding the reaction mixture preheated to 100°C against macroporous silicon under ambient conditions for 12 h could not be detached from the silicon molds because the cohesion between the phenolic resin particles was too low, whereas the phenolic resin particles adhered to the surface of the silicon mold (Figure S5b). Without the presence of $ZnCl_2$ and P123, non-destructive detachment of the phenolic resin from the macroporous silicon molds was likewise impossible. We rationalize this observation as follows. The surface of the macroporous silicon consists of a native silica layer terminated with hydroxyl groups. The $ZnCl_2$ and the PEO block segments of the P123 selectively segregate to the hydroxyl-terminated surfaces of the macroporous silicon and form a wetting layer in contact with the native silica layer. This wetting layer is degraded during sulfuric acid treatment[45] so that the remaining granular phenolic resin stamps can easily be stripped off with tweezers.

The porous phenolic resin stamps were topographically patterned with arrays of contact elements, which were negative replicas of the macropore arrays in the macroporous silicon molds. The height of the contact elements amounted to ≈0.8 μm and matched the depth of the silicon macropores. Because of slight shrinkage of the phenolic resin during the treatment with sulfuric acid, the diameter of the contact elements of ≈0.9 μm was slightly smaller than that of the macropores of the macroporous silicon molds (Figure 1a, b). The contact elements formed monolithic units with a ≈13 μm thick phenolic resin support layer (Figure 1c and d, Figure S6). The porous phenolic resin stamps were penetrated by spongy continuous mesopore networks open to the environment.

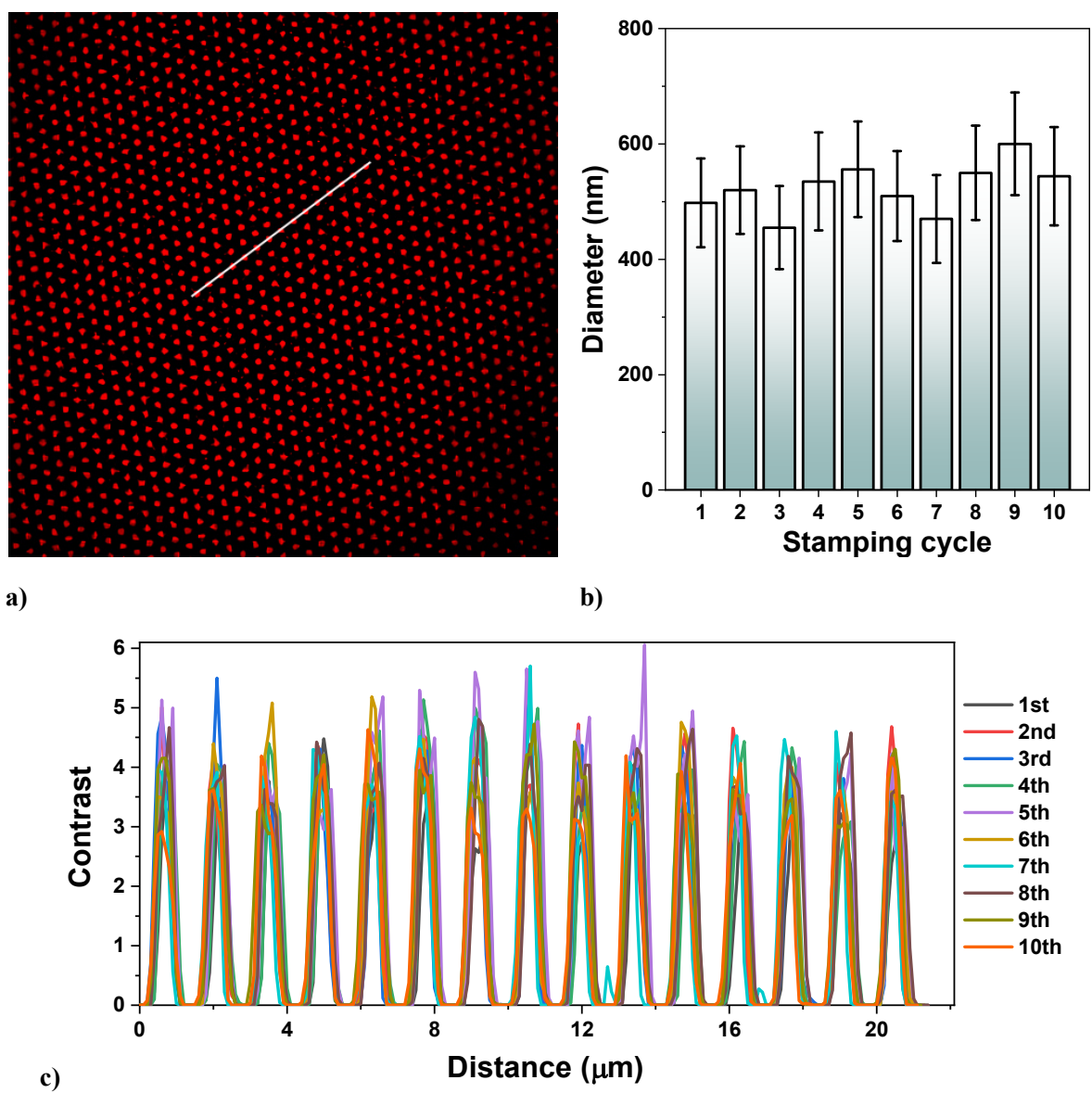

**a)**                    **b)**

**c)**

**Figure 2.** Capillary stamping of ethanolic RhB solutions with phenolic resin stamps. a) Deconvolved total internal reflection fluorescence microscopy (TIRFM) image of RhB microdots on a glass slide obtained by the first stamping step. The displayed image field extends 53.91 μm × 53.91 μm. b) Mean diameters and standard deviations of RhB microdots deposited in 10 successive capillary stamping cycles carried out without reinking of the porous phenolic stamp. At least 150 RhB microdots per TIRFM image were evaluated. c) Linear fluorescence contrast profiles obtained from deconvolved TIRFM images of RhB microdot arrays on glass slides prepared in 10 consecutive stamping steps with the same phenolic resin stamp without reinking. The linear intensity contrast profile obtained from the RhB dot array prepared by the first stamping step is indicated by a white line in panel a).

**2.2. Capillary microstamping with phenolic resin stamps.** We tested capillary stamping as a first application of phenolic resin stamps using the dye Rhodamine B (RhB) as non-volatile model ink component because RhB micropatterns deposited on glass slides are easy to image by total internal reflection fluorescence microscopy (TIRFM). 10 µL of an ethanolic 0.05 wt-% RhB solution was dropped onto the topographically patterned contact surfaces of porous phenolic resin stamps extending 0.5 cm × 1.0 cm. After a wait time of 1 min, residual solution was removed with filter paper. The inked porous phenolic resin stamps were then instantly brought into contact with glass slides for ≈1 s under a pressure of 5.3 kN m². The array of fluorescent RhB microdots obtained in this way was imaged by TIRFM. Since the size of the RhB microdots was close to the diffraction-limited resolution of the TIRF microscope, we deconvolved the TIRFM images by applying the theoretical point spread function (PSF) for wide field microscopes. As revealed by the evaluation of the deconvolved image displayed in Figure 2a, the RhB microdots had a mean diameter of 500 nm ± 80 nm (Figure 2b). The lattice constant of the RhB microdot arrays amounted to ≈1.5 µm corresponding to the lattice constant of the contact element arrays on the porous phenolic resin stamps. We carried out 10 successive stamping cycles using the same phenolic resin stamp without replenishment of ink. No deterioration of the pattern quality was apparent even after the tenth stamping cycle (Figure 2a; Figure S7). The mean RhB microdot diameters scattered about 500 nm (Figures 2b and S8). The fluorescence contrast between the RhB microdots and their surroundings amounted to about 5, as apparent from linear fluorescence contrast profiles taken along lines intersecting 15 RhB microdots. Quantitative comparisons of fluorescence intensities within a TIRFM image or between different TIRFM images are delicate. First, the significance of apparent fluorescence intensity differences is not calibrated and can, therefore, not be assessed. It is, secondly, inevitable that the microscopic slices are slightly misaligned. Differing degrees of misalignment result in differences in the fluorescence intensities. Thirdly, the image field is not evenly illuminated so that fluorescence intensity profiles taken at slightly different positions – either in the same image or in different images – may differ. A fourth effect to be considered is possible fluorescence quenching by dimerization of the dye RhB used here.[46] It should finally be noted that contrast variations apparent in the fluorescence contrast profiles displayed in Figure 2c may result from the fact that the lines, along which we took the fluorescence contrast profiles, do not always cut through the fluorescence maxima of the considered microdots. The evaluation of the 10 TIRFM images (Figure 2a; Figure S7) of RhB microdot arrays, which were successively stamped without replenishment of ink, indicates that – similar to the case of mesoporous silica stamps[47] – the porous phenolic resin stamps contain continuous pore systems. These continuous pore systems can be soaked with ink and act as ink reservoir during the performance of successive stamping steps. It is, in principle, possible to supply fresh ink to the pore system without interruption of ongoing stamping operations.

**2.3. Decal transfer printing on QCM resonators.** Continuous phenolic resin coatings directly synthesized on QCM resonators show poor adhesion and high brittleness. They delaminated in the course of the sulfuric acid treatment to remove $ZnCl_2$ and poloxamer for porosity enhancement. Therefore, we investigated decal transfer printing of phenolic resin microdot arrays using topographically patterned phenolic resin stamps as alternative way to functionalize QCM resonators with phenolic resin. So far, decal transfer printing was typically carried out with solid elastomeric poly(dimethylsiloxane) (PDMS) stamps. It was stressed that solid elastomeric PDMS is far from being optimal for decal transfer since it is a low-modulus material that is considerably stretched prior to rupture.[18] As alternative, decal stamping with spongy nanoporous block copolymer stamps has recently been studied.[21, 48] However, the obtained PDMS and block copolymer microstructures are difficult to modify, as required to introduce further application-specific functionalities. By contrast, micropatterns of materials characterized by high densities of functional groups, like phenolic resins, enable further functionalization of the deposited patterns.

To demonstrate decal transfer printing of phenolic resin microdots, we sputter-coated gold-coated QCM resonators (Figure S9a) with a ≈7 nm thick titanium film (Figure S9b). The surface of the titanium film was converted to titania under ambient conditions within 24 h.[49] The surfaces of the porous phenolic resin stamps and the titania surfaces of the QCM resonators were further activated by oxygen plasma treatment. The contact elements of the porous phenolic resin stamps were thus sharpened (Figure 3a and Figure S10) while their height was reduced to ≈660 nm and their edge length to ≈620 nm. The thus-treated porous phenolic resin stamps and the likewise activated titania-coated QCM resonators were instantly brought into contact for 10 s under a pressure of 2368 kN m$^{-2}$ (Figure 3b). After retraction of the phenolic resin stamps, corrugated phenolic resin microdots remained on the surfaces of the titania-coated QCM resonators (Figure 3c-e). The arrays of the phenolic resin microdots with a lattice constant of 1.5 µm were uniform and nearly free of defects over large areas (Figure S11a). The phenolic resin microdots had edge lengths of ≈770 ± 120 nm (Figure S11b) and heights 10 nm ± 3 nm, as revealed by AFM analysis (Figure 3f and g, Figure S11c). The phenolic resin microdots deposited by decal transfer printing provide a rough, corrugated surface characterized by abundant amounts of exposed hydroxyl groups. Therefore, the phenolic resin microdots are ideally suitable as generic anchor sites for further functionalization specific to the envisioned applications. It should be noted that three-dimensional porosity was not an intended design feature of the phenolic resin microdots. For surface-sensitive sensing methods such as QCM sensing or ATR- and SPR-based sensing methods, a 3D porous material is not the configuration of choice because the analyte molecules need to be captured close to the sensor interface.

After stamping, compressed ruptured contact elements with edge lengths of ≈900 ± 105 nm remained on the surfaces of the phenolic resin stamps, which nevertheless retained their porous morphology (Figure S12a and b). As revealed by AFM topography imaging, the height of the compressed contact elements decreased to ≈150 nm (Figure S12c and d). The added heights of the phenolic resin microdots deposited on the titania-coated QCM resonators and the ruptured contact elements remaining on the porous phenolic resin stamps amount to ≈160 nm, as compared to ≈660 nm prior to the stamping, indicating significant compression. The main drawback of decal transfer printing with porous phenolic resin stamps is, therefore, that the porous phenolic resin stamps cannot be reused.

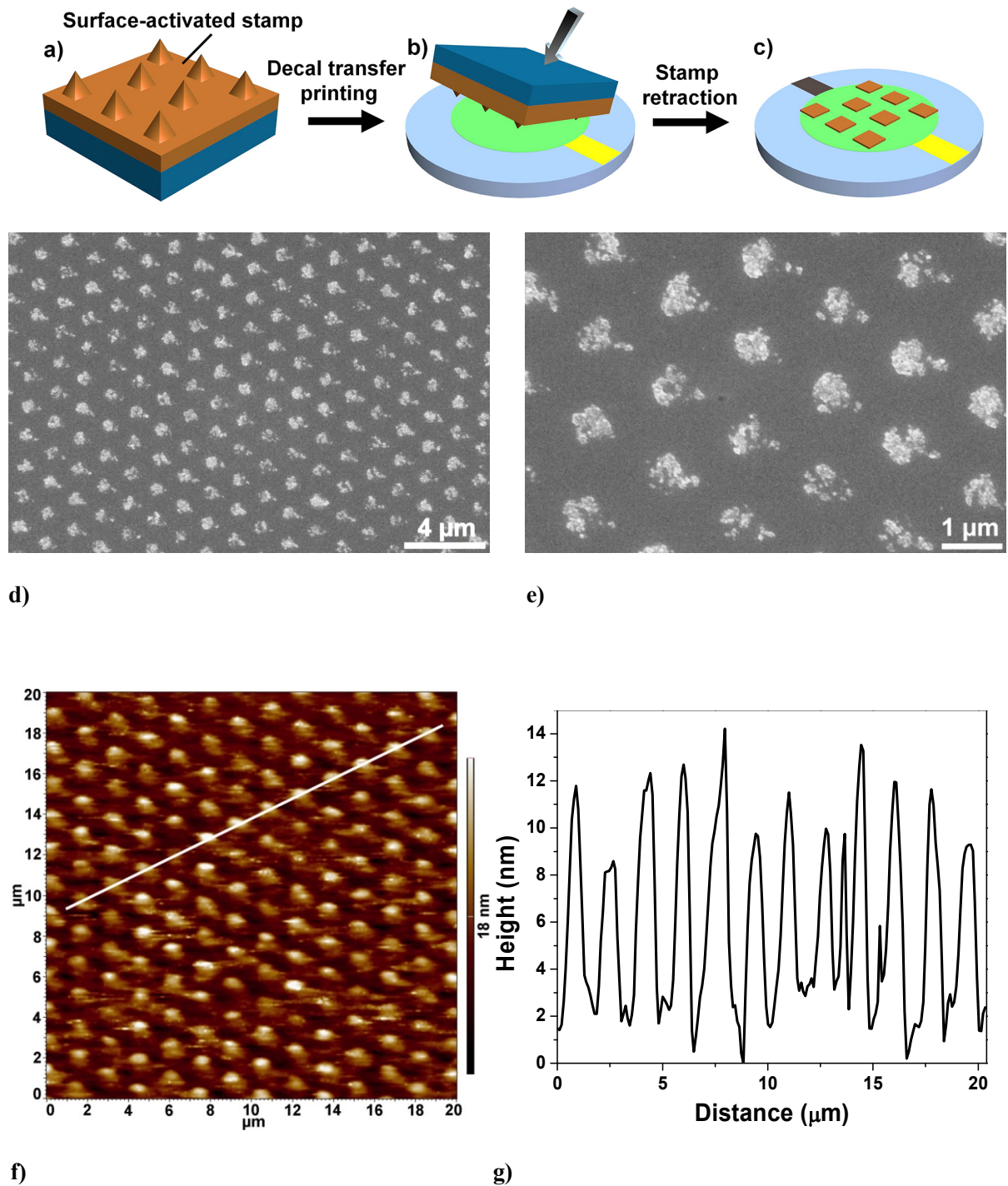

**Figure 3.** Corrugated phenolic resin microdot arrays on QCM resonators. a-c) Fabrication. a) A phenolic resin stamp activated with oxygen plasma is b) brought into contact with the QCM resonator for 10 s under a pressure of 2368 kN m$^{-2}$, yielding c) arrays of corrugated phenolic resin microdots after the retraction of the stamp. d, e) SEM images of arrays of phenolic resin microdots on a QCM resonator. f) AFM topography image of a phenolic resin microdot array on a QCM resonator; g) topographic profile along the white line in panel f).

**2.4. Functionalization of phenolic resin microdots with PEI.** As example for the use of phenolic resin microdots as anchoring sites for further functionalizations, we bound PEI, a polyelectrolyte characterized by abundance of amino groups, to the surfaces of the phenolic resin microdots. We adapted a procedure reported elsewhere[30] and immersed the QCM resonators patterned with phenolic resin microdots into aqueous PEI solutions. PEI molecules adsorb to phenolic resin via electrostatic interactions and the formation of hydrogen bonds (Figure 4a). The functionalization with PEI further boosts the number of functional groups available for interactions with analytes, diversifies the range of available functional groups in that the basic amino groups of the PEI can quickly capture electron pair acceptors and increase the microdot size, that is, the microdot/QCM resonator contact area. As a result, we obtained PEI/phenolic resin hybrid microdots (Figure 4b), which still formed extended ordered arrays (Figure S13a). The edge lengths of the PEI/phenolic resin hybrid microdots amounted to 990 nm ± 140 nm (Figure S13b) – as compared to 770 nm ± 120 nm for the phenolic resin microdots (Figure S11b). The heights of the PEI/phenolic resin hybrid microdots amounted up to 90 nm ± 18 nm (Figure 4c, d and Figure S13c), which is nearly one order of magnitude larger than the initial height of the phenolic resin microdots (Figure 3f, g and Figure S11c). The PEI/phenolic resin hybrid microdots covered ≈24 % of the surface area of the QCM resonators (Figure S14a) as compared to ≈10% covered by the phenolic resin microdots (Figure S14b). The PEI/phenolic resin hybrid microdots still had a highly corrugated surface topography with feature sizes ranging from ≈8 to ≈70 nm (Figure 4b, Figure S15), which is advantageous for sensing as it is associated with a large active surface area.

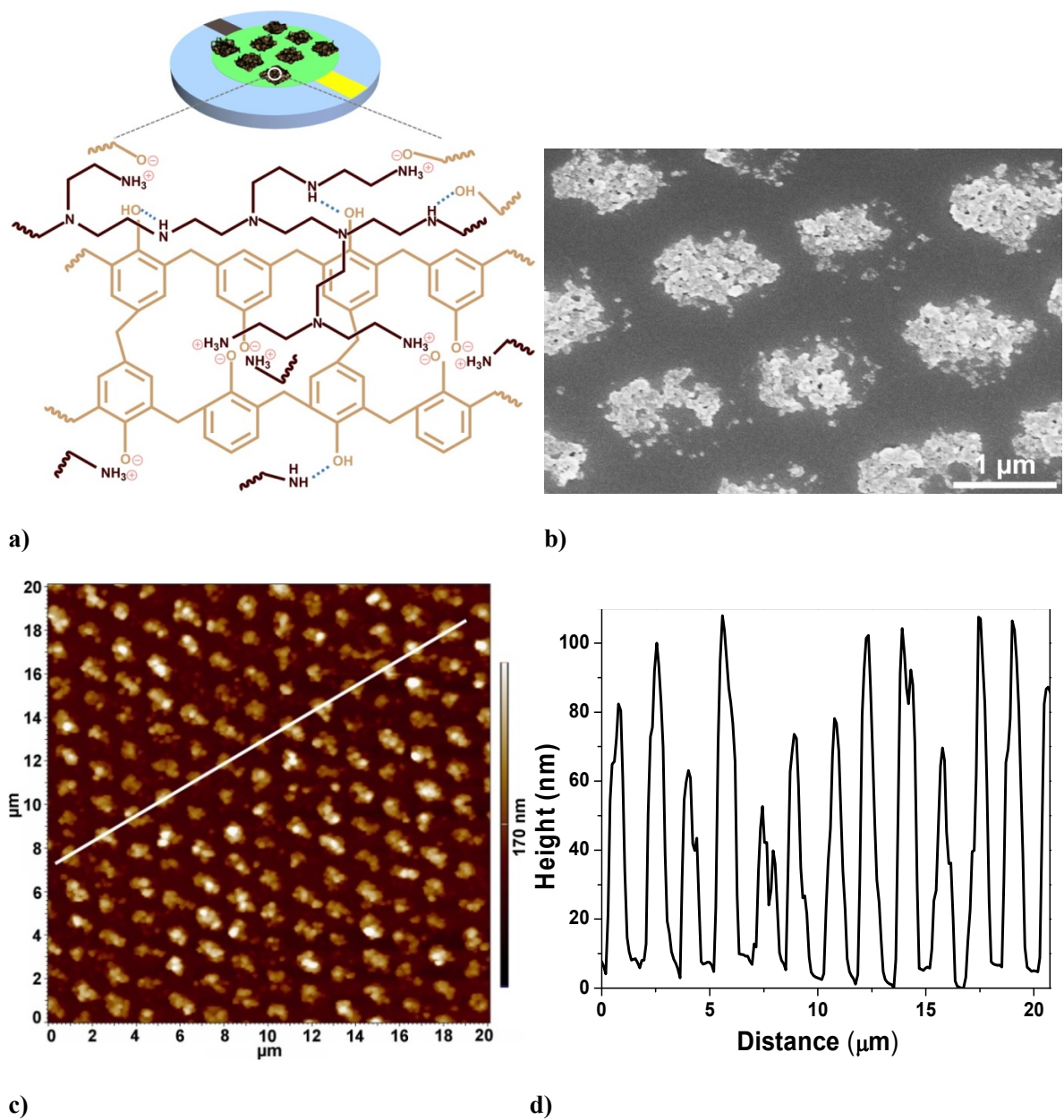

**Figure 4.** Arrays of PEI/phenolic resin hybrid microdots on a QCM resonator. a) Schematic diagram. b) SEM images. c) AFM topography image. d) Topographic profile along the white line in panel c).

The PEI/phenolic resin hybrid microdot arrays on the QCM resonators were further investigated with X-ray photoelectron spectroscopy (XPS; Figure S16). The XPS spectrum of phenolic resin microdot arrays on titania-coated QCM resonators shows O 1s, Ti $2p_{1/2}$, Ti $2p_{3/2}$ and C 1s peaks at binding energies of 531 eV, 464.4 eV, 458.8 eV and 284.8 eV, respectively. The XPS spectrum taken on an array of PEI/phenolic resin hybrid microdots shows essentially the same peaks. However, in addition, a N 1s peak appears at 399 eV that indicates the presence of PEI. In contrast to phenolic resin, PEI does not contain oxygen atoms. Hence, the atomic C/O ratio of 0.7 found for the phenolic resin microdots increased to 1.1 for the PEI/phenolic resin hybrid microdots (Figures S17 and S18, Table S1). Finally, the larger surface coverage by the PEI/phenolic resin hybrid microdots as compared to the phenolic resin microdots decreased the relative surface area where $TiO_2$ was exposed. Thus, the elemental Ti:C and Ti:O ratios were reduced from 1.06 and 0.75 for phenolic resin microdot arrays to 0.24 and 0.26 for arrays of PEI/phenolic resin hybrid microdots. The smaller decrease in the elemental Ti:O ratio as compared to the decrease in the Ti:C ratio is consistent with the assumption that phenolic oxygen atoms were – at least partially – screened by PEI molecules.

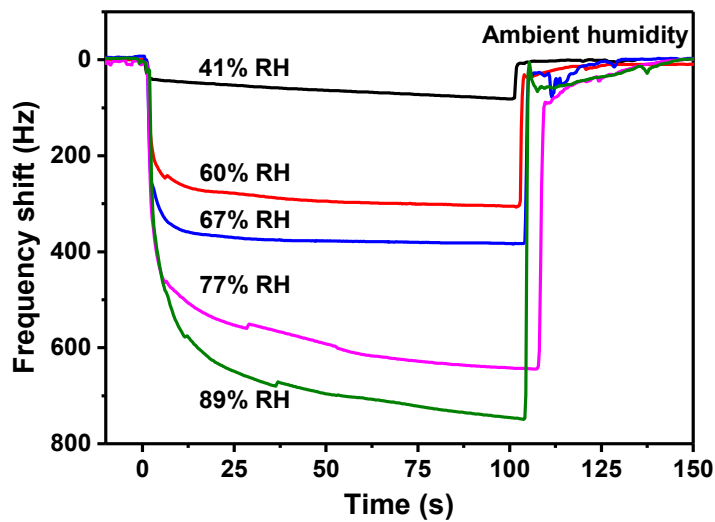

a)

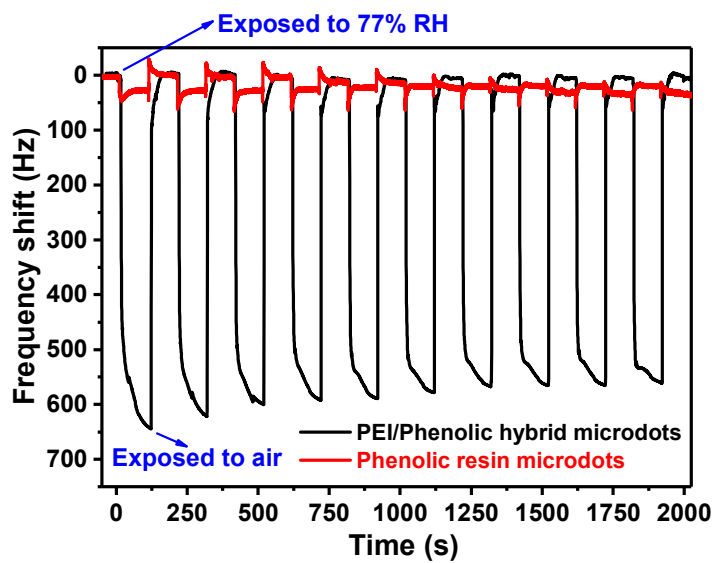

b)

**Figure 5.** Humidity sensing with a QCM resonator functionalized with arrays of PEI/phenolic resin hybrid microdots. a) Frequency shifts observed when the relative humities were increased from a starting value of 29 % to 41 %, 60 %, 67 %, 77 % and 89 %. The resonance frequency shift at the ambient relative humidity of 29 % is set to 0 Hz. b) Frequency shifts obtained by alternating exposure to relative humitities of 29 % and 77 %. Frequency shifts obtained with a QCM resonator functionalized with phenolic resin microdots are shown for comparison.

**2.5 QCM measurements.** QCM resonators are used as preconcentration sensors because mass changes on their surfaces result in detectable changes of their resonant frequencies. To preconcentrate analytes from the gas phase, the surfaces of such QCM resonators need to be functionalized with preconcentration coatings capturing the analytes.[32-35] As first model application for QCM resonators functionalized with PEI/phenolic resin hybrid microdots we studied the detection of varying relative humidities (RH). A QCM resonator functionalized with PEI/phenolic resin hybrid microdots was equilibrated under ambient conditions at a relative humidity of 29 % and then transferred into a sealed container with relative humidities adjusted to 41 %, 60 %, 67 %, 77 % and 89 % (Figure 5a). The stepwise increase in the relative humidity instantaneously triggered stepwise frequency shifts. After only 3 s dwell time in the sealed container, the frequency shifts amounted to 40 Hz, 225 Hz, 300 Hz, 420 Hz and 445 Hz, respectively. After 100 s dwell time, frequency shifts of 80 Hz, 300 Hz, 380 Hz, 640 Hz and 750 Hz were observed. We plotted the frequency shifts obtained at different dwell times in the sealed container against the relative humidity (Figure S19). Already for a dwell time of 5 s the slope of the linear fit amounted to 8.4 Hz/%. For dwell times of 10 s, 20 s, 50 s and 100 s the slopes of the linear fits amounted to 10.1 Hz/%, 11.3 Hz/%, 12.3 Hz/% and 13 Hz/%. After removal from the sealed container and exposure to ambient conditions, the resonance frequency of the QCM resonators relaxed to its initial value in less then 40 s (Figure 5a). To test the reproducibility of the humidity sensing, we subjected a QCM resonator functionalized with PEI/phenolic resin hybrid microdots 10 times to alternating relative humidities of 29 % (ambient conditions) and 77% (in the sealed container) (Figure 5b). The mean frequency shift after 101 s dwell time amounted to 589 ± 58 Hz. A QCM resonator functionalized with as-stamped phenolic resin microdots tested for comparison under the same conditions showed a mean frequency shift of 46 ± 10 Hz, suggesting that the possibility to further functionalize as-stamped phenolic resin microdot arrays with PEI is crucial to achieve the desired functionality.

We selected gaseous formic acid as an acidic model analyte because only a limited number of QCM-based sensor configurations were specifically designed for the detection of formic acid from the gas phase.[50-52] QCM resonators functionalized with arrays PEI/phenolic resin hybrid microdots were placed in sealed containers with a volume of 100 mL. In a first series of experiments, after equilibration for at least 20 minutes liquid formic acid was injected through a septum cap in such a way that the formic acid reservoir was located about 1.5 cm away from the QCM resonator (Figure 6a and b). The initial frequency change rates amounted to about 7 Hz/s for an applied formic acid volume of 0.001 mL, to about 20 Hz/s for an applied formic acid volume of 0.01 mL, to about 19 Hz/s for an applied formic acid volume of 0.1 mL, to about 23 Hz/s for an applied formic acid volume of 1 mL (Figure 6a) and to about 41 Hz/s for an applied formic acid volume of 5 mL (Figure 6b). After an exposure time of 30 s, frequency shifts of 115 Hz, 310 Hz, 317 Hz, 475 Hz and 1000 Hz were detected for applied formic acid volumes of 0.001 mL, 0.01 mL, 0.1 mL, 1 mL and 5 mL. After an exposure time of 600 s, the frequency shifts amounted to 140 Hz for a formic acid volume of 0.001 mL, to 915 Hz for a formic acid volume of 0.01 mL, to 1300 Hz for a formic acid volume of 0.1 mL, to 1870 Hz for a formic acid volume of 1 mL and to 2249 Hz for a formic acid volume of 5 mL (Figure S20). Quantitative comparisons of the capability to preconcentrate formic acid molecules at the surfaces of QCM resonators between the PEI/phenolic resin hybrid microdot arrays studied here and previously reported coatings[50-52] are difficult because experimental designs and reported benchmark parameters significantly vary. For example, García-Verdugo-Caso et al. coated QCM resonators with polyoxyethylene bis(amine) (layer thickness and molecular weight were not disclosed). Mean frequency shifts of 35.7 Hz ± 1.6 Hz and 18.2 Hz ± 1.2 Hz were obtained for formic acid vapor concentrations of 55.0 mg/m$^3$ and 30.0 mg/m$^3$.[50] Yan et al. electrochemically immobilized polyaniline films on the surfaces of QCM resonators.[51] Formic acid vapor concentrations of 1, 5, 10, and 20 mg/m$^3$ resulted in final frequency shifts of 3.4, 6.5, 11.6, and 20 Hz. Lawrence et al.[52] reported a frequency shift of 12312 Hz for microstructured films consisting of the protein cytochrome c, which were prepared by soft-templating with polystyrene spheres. The experimental design we applied is rather comparable with that employed by Lawrence et al.[52] than with the designs applied by García-Verdugo-Caso et al.[50] and Yan et al.[51] In the latter two cases gas mixtures with adjusted formic acid concentrations were pulsed into the test chamber, and the QCM resonators were coated with continuous polymer films. In the work by Lawrence et al. as well as in this work, liquid formic acid was injected into the test chambers (Lawrence et al.: 10 mL into a test chamber with undisclosed volume), and the QCM resonators were coated with micropatterned preconcentration layers. Notably, Lawrence at all found that the frequency shifts obtained with micropatterned cytochrome c films were about twice as high as the frequency shifts obtained with smooth reference samples.[52]

A closer inspection of the development of the frequency shift over time after injection of 5 mL formic acid reveals details regarding the response behavior of QCM resonators functionalized with PEI/phenolic resin hybrid microdots to formic acid vapor (Figure 6b). The steep initial change in the frequency shift by 1300 Hz within 52 s is likely related to the adsorption of formic acid molecules at exposed PEI amino groups (regime 1). A second regime with a duration of ≈145 s characterized by a less steep change in the frequency shift from 1300 Hz to 1930 Hz (regime 2) likely involves the capture of formic acid molecules by PEI amine groups and possibly hydroxy groups of the phenolic resin away from the surface of the PEI/phenolic resin hybrid microdots. Subsequently, the frequency shift further increases slowly from 1930 Hz to 2890 Hz within 1300 s (regime 3), possibly because formic acid molecules associate to secondary and tertiary amino groups of the PEI and to hydroxy groups of the phenolic resin, to which access is impeded by steric hindrance. Similar changes in the steepness of the frequency shift profiles were found for microstructured cytochrome c films and explained by the adsorption of analyte molecules to exposed or buried functional groups of cytochrome c.[52] After the removal of the QCM resonator functionalized with PEI/phenolic resin hybrid microdots from the sealed container the resonance frequency approached its initial value prior to the exposure to formic acid within a few seconds, indicating rapid desorption of the formic acid molecules. The QCM resonator functionalized with PEI/phenolic resin hybrid microdots was subjected nine more times to the sensing cycle shown in Figure 6a. The frequency shifts obtained after exposure to formic acid for 1500 s remained by and large constant and amounted to 2924 ± 64 Hz (Figure 6c). Such as for the sensing of changes in the relative humidity, functionalization of the as-stamped phenolic resin microdots with PEI is crucial for efficient formic acid preconcentration sensing. For example, the initial frequency change rate right after the injection of 5 mL formic acid decreased from 41.2 Hz/s for QCM resonators functionalized with PEI/phenolic hybrid dots to 7.3 Hz/s for QCM resonators with as-stamped phenolic resin microdots (Figure 6d).

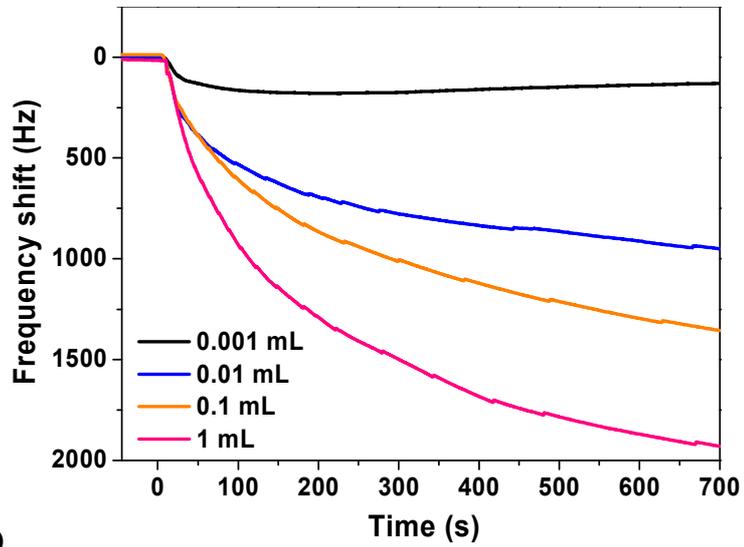

a)

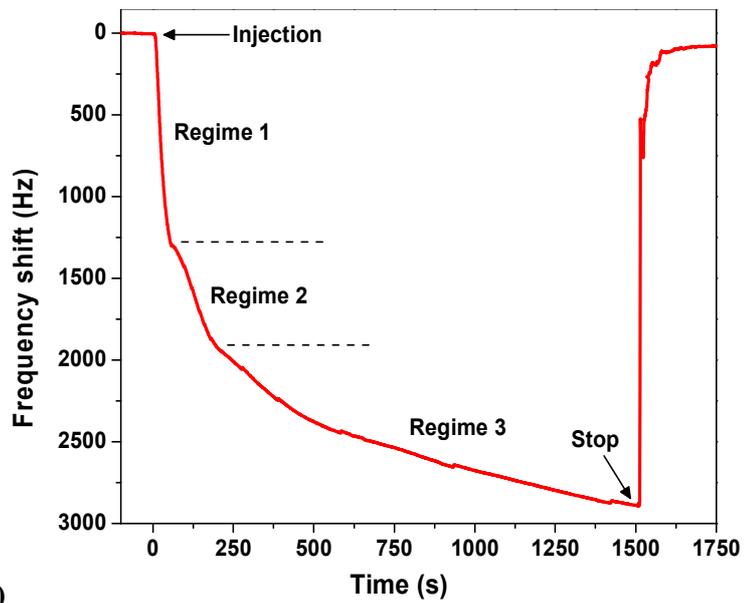

b)

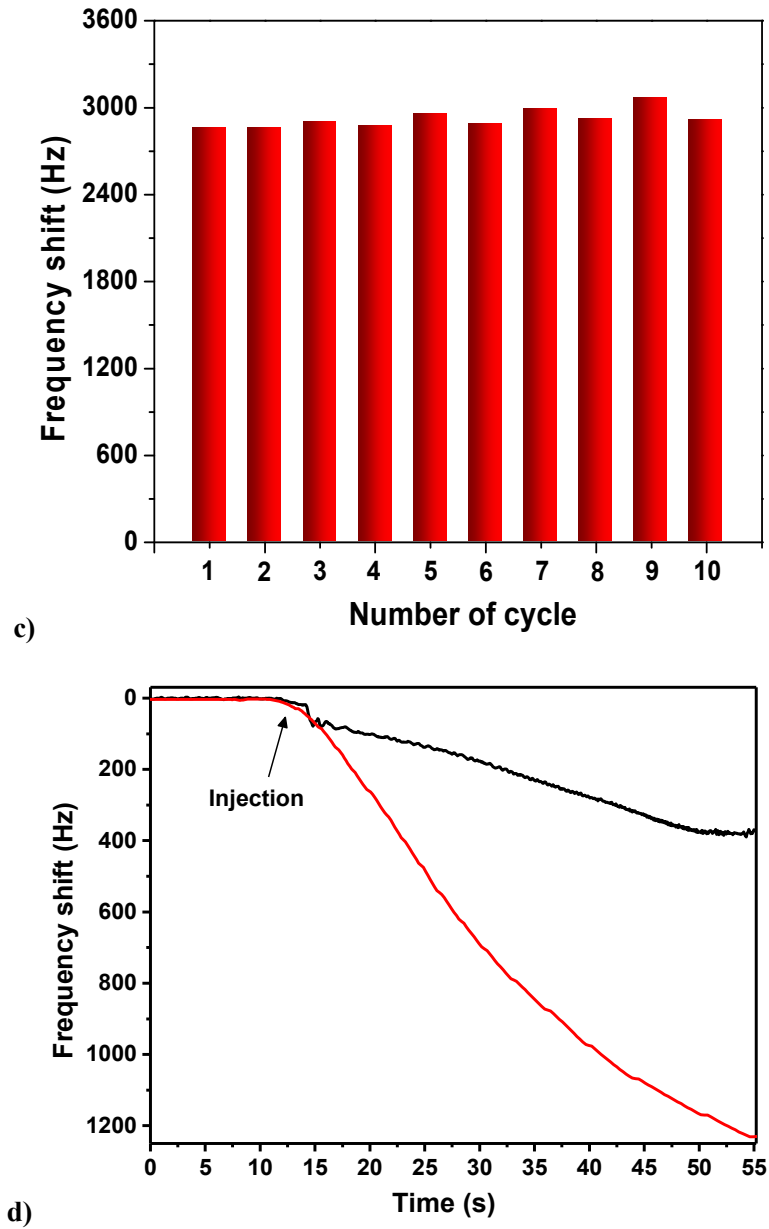

c)

d)

**Figure 6.** Preconcentration sensing of formic acid vapors with QCM resonators functionalized with PEI/phenolic hybrid microdot arrays located in sealed containers. a) Frequency shifts versus exposure time after injection of 0.001 mL, 0.01 mL, 0.1 mL and 1 mL formic acid. b) Frequency shift versus exposure time after injection of 5 mL formic acid. c) Frequency shifts in successive sensing cycles involving equilibration under ambient conditions followed by exposure to gaseous formic acid for 1500 s. d) Comparison the preconcentration sensing of formic acid vapors with QCM resonators functionalized with phenolic resin microdot arrays (black curve) and PEI/phenolic resin hybrid microdot arrays (red curve). The frequency shift is plotted against the exposure time after injection of 5 mL formic acid.

## 3. Conclusions

Additive parallel substrate micropatterning by means of stamps may involve nondestructive lithographic ink transfer under moderate contact pressure or lithographic decal transfer of parts of the stamps under high contact pressure. We devised an optimized two-step thermopolymerization combined with replication molding that yielded granular-porous dual-use phenolic resin stamps with topographically micropatterned contact surfaces. These phenolic resin stamps enable capillary stamping by exploiting the interstices between the phenolic resin nanoparticles forming the stamps for ink storage and ink supply. Thus, multiple stamping steps yielding micropatterns on glass substrates were manually performed under ambient conditions. No loss of micropattern quality was evident, and no interruptions for ink replenishment were necessary. The second use of the porous phenolic resin stamps involved the transfer of the tips of the stamps' contact elements to counterpart substrates by decal transfer printing under high contact pressure. The porous-granular nature of the phenolic resin stamps facilitated the rupture of the contact elements in the course of stamp retraction. In this way, substrates for preconcentration sensing by, for example, ATR infrared spectroscopy, SPR spectroscopy or QCM sensing may be functionalized with phenolic resin microdot arrays characterized by large numbers of exposed hydroxyl groups, which either act as preconcentration coating themselves or as anchoring sites for further application-specific functionalizations. The deposition of phenolic resin microdot arrays by decal transfer printing has the following advantages. 1) In contrast to other coating methods, good adhesion of the phenolic resin to the substrates is achieved because of the high applied contact pressure. 2) Delamination and wrinkling, which may pose problems in the case of continuous coatings, remain confined to single affected phenolic resin microdots and are stopped at the uncovered area between the discrete phenolic resin microdots. 3) If relatively small phenolic resin microdots act as anchor sites for functional long-chain polymers, such as PEI, the contact areas between the latter and the substrates may be large so that many analyte molecules captured by the functional long-chain polymers are in immediate proximity to the substrate surface. To demonstrate this functionalization algorithm, we deposited phenolic resin microdots by decal transfer printing onto QCM resonators and adsorbed the polymer PEI having a large density of amino groups on the phenolic resin microdots. As application examples, we then demonstrated preconcentration sensing of humidity and gaseous formic acid. The porous phenolic resin stamps presented here may be representative of tailored next-generation multiple-use stamps for microcontact printing with functionalities going beyond those of the established solid-elastomeric PDMS stamps.

# 4. Experimental Section

*Materials:* Branched polyethylenimine (50 wt-% in $H_2O$; $M_n$ = 60000 g/mol; $M_w$ = 750000 g/mol), $ZnCl_2$ ($\geq$ 97%), formic acid (95 % - 97 %), acetic acid ($\geq$ 99.8 %), cyclohexane ($\geq$ 99.5 %), toluene (99.8 %), hydrochloric acid (ACS reagent), phenol ($\geq$ 99.5 %), rhodamine B (RhB; $C_{28}H_{31}ClN_2O_3$; $\geq$ 95 %) and Pluronic P123 ($PEO_{20}$-*b*-$PPO_{70}$-*b*-$PEO_{20}$) triblock copolymer were purchased from Sigma-Aldrich. Propionic acid (ACS reagent) was obtained from Acros Organics. Anhydrous ethanol ($\geq$ 99.997 %) was supplied by VWR. Aqueous formaldehyde solution (37 wt-%) was obtained from Merck. NaOH ($\geq$ 97 %) was purchased from Fluka. $H_2SO_4$ (96 %) was obtained from ORG Laborchemie GmbH. Hexane (95 %) was supplied by Fisher Chemical. All chemicals were used without further purification. Macroporous silicon molds (lattice constant: 1.5 μm, pore depth: $\approx$810 nm, base diameter: $\approx$1 μm; SmartMembranes, Germany, cf. Figure S1) were cut into pieces extending 2.0 cm $\times$ 2.0 cm, ultrasonicated for at least two times in ethanol and dried in nitrogen before use.

*Preparation of resol solutions:* Resol was synthesized by base-catalyzed polymerization of phenol and formaldehyde following protocols described elsewhere.[23, 30, 31] In a typical synthesis procedure, 0.610 g (6.5 mmol) phenol was at first melted at 45 °C. Then, 0.130 g of a 20 wt-% aqueous NaOH solution was added under stirring. The stirring was continued for 10 more minutes. Subsequently, 1.054 g of a 37 wt-% aqueous formaldehyde solution was added dropwise. After stirring at 75 °C for 1 h, the reaction mixture was cooled to room temperature followed by adjustment of the pH value to 7.0 with 0.6 M HCl. The mixture was dried under vacuum at 45 °C for 24 h to remove water. The as-prepared resol was dissolved in 5 g ethanol. The obtained ethanolic resol solution was then filtered with syringe filters having a nominal pore diameter of 0.45 μm to remove NaCl precipitates. 0.678 g P123 dissolved in 10 g ethanol was added to the filtered ethanolic resol solution, and the obtained mixture was stirred for at least 12 h at room temperature. The molar ratio P123 to phenol amounted to 0.018; the integral mass fraction of the non-volatile solutes resol and P123 amounted to 10 wt-%.

*Preparation of porous phenolic resin stamps:* The ethanolic solution containing resol and P123 prepared as described above was diluted with ethanol to reduce the mass fraction of the nonvolatile solutes resol and P123 to 5 wt-%. Then, 0.05 g (0.00037 mol) $ZnCl_2$ was added to an aliquot of 1 g of the diluted resol/P123 solution so that the molar ratio of $ZnCl_2$ to P123 amounted to ≈100. 180 μL of the obtained solution was drop-cast onto macroporous silicon molds with edge lengths of 2 cm and then instantly transferred into an oven preheated to 40 °C for gelation in air. After 2 h, the temperature of the oven was increased to 100 °C at a rate of 1 °C/min. The samples were kept at this temperature for 12 h. The formed phenolic resin stamps attached to macroporous silicon were treated with 55 wt-% $H_2SO_4$ at 100°C for 4 h to remove $ZnCl_2$ as well as P123, copiously rinsed with and then immersed into water. After peeling off the phenolic resin stamps from the macroporous silicon molds under water with tweezers, the detached phenolic resin layers were fished up with silicon wafer pieces extending 2.0 cm × 2.0 cm. After drying under ambient conditions, the phenolic resin layers tightly adhered to the underlying silicon wafer pieces.

*Capillary stamping:* The phenolic resin stamps along with the underlying silicon wafer pieces were cut into pieces extending 0.5 cm × 1.0 cm, which were then glued onto cylindrical stamp holders with a weight of 27 g. To stamp RhB solutions, 10 μL of an ethanolic solution containing 0.05 wt% RhB were cast onto the exposed surfaces of the phenolic resin stamps. After 1 min, residual solution was removed with filter paper. Then, capillary microstamping was manually carried out under ambient conditions with a contact time of ≈1 s and by applying a pressure of 5.3 kN m$^{-2}$.

*Functionalization of QCM resonators by decal transfer printing:* Gold-coated QCM resonators (5 MHz AT-cut quartz with an overlap area of 0.196 cm$^2$ and a thickness of ≈0.333 mm supplied by Quarztechnik

Daun GmbH) were sputter-coated with a ≈7 nm thick Ti layer using a Balzers BAE 120 evaporator applying recipes reported elsewhere.[14] The Ti-coated QCM resonators were exposed to air under ambient conditions (24 °C, 50% relative humidity) for 24 h to form ≈2 nm thick $TiO_2$ layers on the Ti coatings following a procedure reported previously.[47] The $TiO_2$/Ti-coated QCM resonators and the phenolic resin stamps were treated with oxygen plasma (Diener electronic, Ebhausen, Germany) at 100 W for 5 min. The phenolic resin stamps were then cut into pieces extending 0.5 cm × 0.5 cm and glued onto cylindrical stamp holders with a mass of 40 g using double-sided adhesive tape in such a way that the nanorods of the phenolic resin stamps were exposed. Decal transfer printing was carried out for ≈10 s under a pressure of ≈2368 kN m$^{-2}$. To functionalize as-stamped phenolic resin microdot arrays with PEI, QCM resonators functionalized with phenolic resin microdot arrays were immersed into a mixture of 24 mL deionized water and 6 mL PEI solution (≈10 wt-% PEI; cf. subsection "Materials" above) for 30 min at 24 °C. Then, the functionalized QCM resonators were rinsed with deionized water several times.

*QCM sensing:* All measurements were carried out with an eQCM 10M device manufactured by Gamry Instruments. Frequency shifts related to variations of the relative humidity were measured by transferring the QCM resonators into or by removing the QCM resonators from a desiccator. The relative humidity inside the desiccator was adjusted to the specified values by means of saturated aqueous $KNO_3$ solutions. The QCM resonator was exposed to ambient conditions (relative humidity 29 %, temperature: 26°C) until the frequency shift was stable (at least 20 minutes) and then transferred into the desiccator. The desiccator was then immediately sealed. To terminate a QCM measurement, the QCM resonator was removed from the desiccator and exposed to ambient conditions. Frequency shifts triggered by the presence of liquid formic acid reservoirs were measured at 23 °C at an ambient relative humidity of 29%. The QCM resonators were located in a sealed tailormade three-necked flask with a volume of 100 mL and equilibrated for at least for 20 min until a stable response was reached. During the equilibration period, the conditions inside the flask corresponded to ambient conditions outside the flask. Subsequently, the specified volume of liquid formic acid was injected with a syringe through a septum cap in such a way that the liquid was located at the bottom of the flask. The QCM resonator did not come into contact with the liquid; the distance between the QCM resonator and the liquid reservoir amounted to ≈1.5 cm. To terminate a QCM measurement, the QCM resonator was removed from the three-necked flask and exposed to air under ambient conditions. To perform series of successive QCM measurements, the QCM was also equilibrated for at least for 20 min prior to the injection of liquid. After exposure times of 1500 s to the analyte vapor, the QCM resonator was taken out from the sealed flask and exposed to air under ambient conditions for 2 h prior to the next QCM measurement that was carried out as described above.

*Scanning electron microscopy:* Scanning electron microscopy was carried out using a Zeiss Auriga device operated at an accelerating voltage of 7 kV. Before the SEM investigations, the samples were sputter-coated with a ≈5 nm thick iridium layer.[11] The software *Nano Measurer* (developed by the Department of Chemistry, Fudan University, China) was employed to estimate the mean edge lengths and the corresponding standard deviations of phenolic resin microdots as well as of PEI/phenolic resin hybrid microdots by the evaluation of SEM images considering >250 microdots in >3 randomly selected different areas. The fractions of the QCM resonator surfaces covered by phenolic resin microdots or by PEI/phenolic hybrid microdots were estimated from the binarized SEM images shown in Supporting Information Figure S14 using the software *ImageJ*.

*TIRFM microscopy:* TIRF microscopy was performed under ambient conditions following procedures reported by Philippi et al.[47] using an inverted microscope Olympus IX71 equipped with a 4-Line TIRF condenser Olympus TIRF 4-Line. The microscopic images were acquired using a 150-fold oil immersion objective (PLAPON 60x TIRF, NA 1.42) combined with a 561 nm diode-pumped solid-state laser (Cobolt Jive 561), a TIRF pentaband polychroic beamsplitter (Semrock zt405/488/561/640/730rpc) and a pentabandpass emitter (BrightLine HC 440/521/607/694/809). The micrographs were captured with an electron multiplying CCD camera (Andor iXon Ultra 897) and processed using Olympus CellSens Software. Subsequently, the TIRFM images were deconvolved by applying the theoretical point spread function for wide field microscopes using the software Huygens (Huygens Professional; version 20.04; Scientific Volume Imaging B.V., Netherlands; http://svi.nl). The deconvolution parameters were set as follows. Deconvolution algorithm: classic maximum likelihood estimation (CMLE); background mode:

automatic; signal to noise ratio (SNR): 30; number of iterations: 20; quality change number 0.01. The software *Nano Measurer* was used to determine the RhB microdot diameters by evaluation of >150 dots per TIRFM image. The fluorescent contrast was determined as follows. 1) Line scans were taken from deconvoluted images. 2) The background fluorescence intensity outside the RhB microdots was subtracted. 3) The arithmetic mean value of the background-corrected fluorescence minima was calculated. 4) The fluorescence values were divided by the mean fluorescence value of the fluorescence minima.

*Atomic force microscopy.* AFM was carried out using a NTEGRA microscope (NT-MDT) in the tapping mode. The tip curvature radius was < 10 nm. The cantilevers had a length of $94 \pm 2$ μm, a width of $34 \pm 3$ μm and a thickness of $1.85 \pm 0.15$ μm. The force constant and the resonant frequency amounted to 12 N/m ($\pm$ 20 %) and 235 kHz ($\pm$ 10 %), respectively. Histograms of the heights of phenolic resin microdots before and after PEI deposition were obtained by evaluation of >100 microdots in >3 randomly selected different areas in AFM topography images.

*X-ray photoelectron spectroscopy.* XPS measurements were carried out using an ESCA-unit Phi 5000 VersaProbe III with a monochromatic Al Kα (1486.6 eV) X-ray source and an electrostatic hemispherical electronic analyzer (overall resolution 0.4 eV). The take-off angle was set to 45° for all measurements. The spectra were calibrated to the C 1s line of aliphatic carbon (284.8 eV). Atomic ratios (Table S1) were calculated from the narrow-scan XPS spectra shown in Figure S17 and S18 by comparison of normalized peak areas. First, linear (C 1s, N 1s, O 1s) and Shirley (Ti 2p) baseline corrections were performed with the software XPSPEAK 4.1. The baseline-corrected peak areas were subsequently normalized using elemental sensitivity factors of 1.0, 1.8, 2.93, 2.69 and 5.22 for the C 1s, N 1s, O 1s, Ti $2p_{1/2}$ and Ti $2p_{3/2}$ peaks.[53]

ASSOCIATED CONTENT

**Supporting Information**

Figure S1: SEM images of a macroporous silicon mold; Figure S2: scheme of replication molding; Figure S3: scheme of the thermopolymerization; Figure S4: SEM image of macroporous silicon after lift-off; Figure S5: one-step thermopolymerization: SEM images of a phenolic resin stamp and macroporous silicon after lift-off; Figure S6: SEM images of a porous phenolic resin stamp; Figure S7: TIRFM images of stamped RhB microdot arrays; Figure S8: diameter histograms of RhB microdots; Figure S9: scheme of a QCM resonator; Figure S10: SEM images of a phenolic resin stamp after oxygen plasma treatment; Figure S11: QCM resonator patterned with phenolic resin microdots (SEM image, histograms); Figure S12: phenolic resin stamps after decal transfer printing (SEM and AFM); Figure S13: QCM resonator patterned with PEI/phenolic resin hybrid microdots (SEM image, histograms); Figure S14: binarized SEM images of microdot arrays; Figure S15: high-resolution SEM image of a PEI/phenolic resin hybrid microdot; Figures S16-S18 and Table S1: XPS analyses; Figures S19 and S20: additional evaluations of QCM measurements.

AUTHOR INFORMATION

**Corresponding Authors**

*Leiming Guo; E-mail: leiming.guo@uos.de

*Martin Steinhart; E-mail: martin.steinhart@uos.de

**Author Contributions**

The manuscript was written through contributions of all authors. All authors have given approval to the final version of the manuscript.

**Notes**

The authors declare no competing financial interest.

ACKNOWLEDGMENT

The authors thank the European Research Council (ERC-CoG-2014, Project 646742 INCANA) for funding. TIRF microscopy and image analysis was supported by the "Integrated Bioimaging Facility Osnabrück" funded by the German Research Foundation (PI 405/14-1).

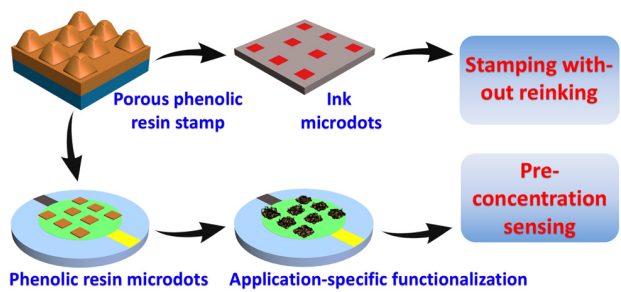

TOC Figure